MEMRISTOR BEHAVIOUR in NANO-SIZED VERTICAL LSMO/LSMO TUNNEL JUNCTIONS

V. Moshnyaga[1], M. Esseling[1], L. Sudheendra[1], O.I. Lebedev[2], K. Gehrke[1], G. Van Tendeloo[2], K. Samwer[1]

[1] *Erstes Physikalisches Institut, Universität Göttingen, Friedrich-Hund-Platz 1, 37077 Göttingen, Germany*
[2] *EMAT, University of Antwerp, Groenenborgerlaan 171, B-2020 Antwerpen, Belgium*

We report a memory resistance (memristor) behavior with nonlinear current-voltage characteristics and bipolar hysteretic resistance switching in the nanocolumnar $La_{0.7}Sr_{0.3}MnO_3(111)/Al_2O_3(0001)(LSMO)$ films. The switching from a high ($R_{HRS}$~$100\ k\Omega$) to a low ($R_{LRS}$~$20\ k\Omega$) resistance occurs at a bias field, $E_C$~$10^6\ V/cm$. Applied electric field drops mostly at the insulating interfacial *LSMO* layer and couples to correlated polarons at the 30°-misoriented *LSMO(111)/LSMO(111)* vertical interfaces. The observed memristance behaviour has an electronic (polaronic) origin and is caused by an electric-field-controlled Jahn-Teller (JT) effect, followed by the orbital reconstruction and formation of a metastable orbitally disordered interfacial phase (*LRS*). Compared to the reported ionic memristor in $TiO_{2-x}$ films, an electronic (polaronic) nano-sized LSMO memristor shows an additional (re-entrant) *LRS-HRS* switching at higher fields because of the second minimum in the elastic energy of a JT system.

Electric pulsed resistance switching (*EPIR*)[1-9] has a large potential for high density non-volatile *RRAM* applications, based on bi- and multi-stable nanodevices[10]. *EPIR* behaviour, observed in different material systems like oxide[1-7] and organic thin films[8.9], is characterised by the two basic features: 1) the bipolarity of switching, provided that high (*HRS*) and low (*LRS*) resistance states can be obtained at positive and negative voltages, respectively; and 2) the switching hysteresis, i.e. *HRS* and *LRS* states are well defined and reproducible, i.e. they are energetically separated. The origin of *EPIR* is still discussed, but it was argued[6,7] that at least the motion of ions in applied electric field, i.e. electro-migration, may play an important role. According to Ignatiev's group[7], the voltage pulses applied to the interfacial depletion layer close to metallic contacts in perovskite based structures, like Ag/PrCaMnO/Pt, activate oxygen in- and out-transport and lead to the resistance change. Recently Strukov et al[11] have shown that the so called "memristor" concept provides a reasonable phenomenological basis for *EPIR*. The memristor ("memory resistor", *M*) is the proposed earlier by Chua[12] fourth fundamental element of electronic circuits, which describes a coupling, $d\varphi=M(q,w)*dq$, between magnetic flux, $d\varphi$, and charge, $dq$. The proportionality coefficient, i.e. the memristance, $M(q,w)$, depends on the charge and the so called "state variable", $w$. The latter being current dependent, $dw/dt=I$, determines the nonlinear electrical properties of a memristor: $U(t)=M(q,w)*I(t)$. The most interesting is the $M(w)$ dependence, which allows one to realize an M-behaviour experimentally as was recently demonstrated in $TiO_{2-x}$ films[13]. Here, the movement of oxygen vacancies in applied electric field across the film thickness, $D$, causes the $TiO_{2-x} \rightarrow TiO_2$ transformation in a part of the film with thickness, $w$, and yields to the resistance change. Very important for applications is the fact that memristance, $M\sim 1/D^2$, strongly increases at the nanoscale because the modulation depth, $w/D$, increases with decreasing the thickness of the film. Very recently Alexandrov and Bratkovsky[14] have developed theoretically an alternative memristor approach, i.e. a polaronic memristor, realized as a molecular quantum dot, built by correlated polarons connected to metallic leads. This, in contrast to[11], can be viewed as a an electronic memristor, in which the state variable instead of the migration of any charged species is provided by a multilevel energy scheme and electronic (vibronic) transitions. In this letter we report experimental observation of an electronic (polaronic) memristor behaviour in the nanocolumnar $La_{0.7}Sr_{0.3}MnO_3$ *(LSMO)* films with vertical nanosized *LSMO/LSMO* tunnel junctions. The results are discussed considering that the state variable, $w$, is controlled by the two different orbital states (with and without Jahn-Teller (*JT*) distortions) of the interfacial *LSMO*. The proposed model assumes that *JT* distortions of the $MnO_6$ octahedra at the interfaces can be influenced by an external electric field, which couples to correlated polarons at the interface.

*LSMO* films with a thickness of *70* and *50 nm* were grown on $Al_2O_3(0001)$ substrates by a metalorganic aerosol deposition technique[15,16]. Crystal structure was characterized by X-ray diffraction (*XRD*) and the film microstructure by scanning electron microscopy (*SEM*), transmission electron microscopy (*TEM*) and selected-area electron diffraction (*SAED*). Transport measurements were carried out on micro-bridges (see Fig. 1), patterned by electron beam lithography, followed by dry etching. The so called “narrow” bridge had a length, $L=1.25\ \mu$, and a width, $W_1=1.25\ \mu$, while for the “wide” bridge – $L=1.25\ \mu$ and $W_2=2.2\ \mu$. The contacts were made by the evaporation of *Au/Cr* films through a mask. The 4-probe dc resistance was measured in a constant-current mode, $I=1\text{-}2000\ \mu A$, for temperatures, $T=5\text{-}400\ K$, and magnetic fields, $\mu_0 H=0\text{-}5\ T$, in a He-bath cryostat with a superconducting solenoid. Magnetization measurements of the whole film samples (5x10 $mm^2$) were done by using MPMS system (“Quantum Design”) for $T=2\text{-}400\ K$.

The characteristics of a typical nanocolumnar *LSMO* film are presented in Fig. 1. *XRD* shows a predominance (98 %) of an out-of-plane (111)-texture of a pseudo-cubic *LSMO* with a lattice constant, $a=0.386\ nm$. The low-magnification cross-section *TEM* image in the inset to Fig. 1a) illustrates that the film is composed of nanocolumns, growing from the substrate. On a SEM image (Fig. 1c) one can see nanocrystalline blocks (grains) of triangular shape with a mean diameter, $D=40\pm 5\ nm$. The temperature dependences of the resistance and magnetization of the whole film sample (Fig. 1b) show coupled metal-insulator ($T_{MI}$) and ferromagnetic ($T_C$) transitions, $T_C$~$T_{MI}$=370 K, typical for an optimally doped *LSMO*. An increased resistance at low temperatures as well as an insulating-like dependence with $dR/dT>0$ for T<30-40 K (see the inset in Fig. 1b) indicate that the internal interfaces (grain boundaries) govern electron transport in the nanocolumnar *LSMO* films, resulting in a large low-field TMR=34 % (Fig. 1c)).

Both "narrow" and "wide" microbridges show essentially nonlinear electric behaviour, followed by the resistance switching. The resistance of a "narrow" bridge in Fig. 2 a) decreases significantly by increasing the current from $J\sim 1\ \mu A$ up to a threshold current $J_C\sim 120\ \mu A$. By exceeding $J_C$ the resistance drops abruptly, indicating the existence of *HRS* and *LRS*; these states were well separated from each other and reproducible. With a further increase of the current up to a maximum value, $J=200\ \mu A$, no large changes in the resistance were observed. By changing the current direction the *LRS* survives for small currents, and then switches back to the initial *HRS* at $J\sim -115\ \mu A$. The *LRS* in a "wide" bridge is found to be unstable for small currents, "creeping" with decreasing the current back to the *HRS* (not shown). Note that magnetic field does not affect $J_C$ in all studied microbridges, thus ruling out current-induced switching[17] as a possible origin of the observed behaviour. Joule heating can be also excluded because the resistance in *LRS* for large currents, $J>J_C$, was much lower than that for *HRS* ($J<J_C$) for all temperatures, $5\ K<T<400\ K$.

In Fig. 2b) the evaluated current-voltage, *I(U)*, characteristics are shown. Remarkably, despite of the large difference in the values of "critical" current densities, i.e. $J_C=2*10^5\ A/cm^2$ and $J_C=10^6\ A/cm^2$ for the "narrow" and "wide" bridges, respectively, the switching bias voltages were found to be the same, $U_C=8\ V$. A similar value, $U_C=6\ V$, was obtained for an additional bridge with $L=1\mu$ and $W=1\ \mu$, shown in Fig. 2 c), d). Due to relatively small switching amplitude ($HRS=39\ k\Omega$ and $LRS=33\ k\Omega$) the bias voltage drops by switching very moderately, allowing us to follow the "major" switching loop with the second and final switching back to the *HRS* at $J\sim 190\ \mu A$. These experiments demonstrate that the resistance switching is driven by applied electric field and not by the current. To be sure that *LSMO/LSMO* interfaces are necessary for the switching we also measured (see Fig. 2 e)) a microbridge ($L=1.5\ \mu$, $W=2\ \mu$), patterned on an epitaxial *LSMO/MgO(100)* film[18]. The *R(I)* curve in Fig. 2e) demonstrates no signs of resistance switching for $J=0-10\ mA$ or $U=0-4\ V$, but rather Joule heating effect with $R\sim I^2$ behavior. Note, that the absence of switching in epitaxial microbridge rules out the ionic (oxygen in- and out-diffusion at the Au/manganite interface)[7] mechanism of the resistance switching in nanocolumnar LSMO/$Al_2O_3$(0001) films. Moreover, low resistivity, $\rho(\mu_0 H=0)\sim 10^{-3}\ \Omega cm$, and very small $CMR=(R(0)-R(7\ T))/R(0)\sim 1\%$, at $T=5\ K$ (not shown) observed in the "epitaxial" bridge, i.e. typical behaviour of epitaxial LSMO films, evidences that e-beam lithography itself does not deteriorate manganite films. Thus, to observe an electric field induced resistance switching in *LSMO* the nanocolumnar structure is necessary. The insulating barriers where the voltage drops are provided most probably by the vertical *LSMO/LSMO* interfaces, which demonstrate tunneling characteristics. Fig. 2f) shows that measured conductivity, $\sigma(U)=dJ/dU$, for the "narrow" microbridge in *HRS* can be fitted well within the Simmons model[19], which describes a conductivity of a planar tunnelling contact with a barrier thickness, *d*, and height, ϕ: $\sigma(U) = \sigma_0 + \sigma_1(d,\phi)\times U^2$. However, one gets an unphysical barrier height, $\phi=57\ eV$, assuming that only one barrier with $d\sim 1\ nm$ is available. A more realistic assumption, dictated by the nanocolumnar architecture, is a serial connection of about $N=L/D\sim 30$ barriers along the microbridge length. Giving realistic barrier thickness, $d=3\ nm$[20], a reasonable barrier height, $\phi=1.8\ eV$, is obtained. These crude estimations illustrate that the transport in nanocolumnar *LSMO* films is of tunnelling type and one has to consider a network of tunnelling junctions. The observed low-field *TMR=34 %* (see Fig. 1d)) confirms a decisive role of the tunnelling. Note that *TMR=6 %* (Fig. 1e)) was observed even for *LRS*.

In plan-view *TEM/HREM* images, shown in Fig. 3a), one can see the detailed topology of the *LSMO* nanostructure. *SAED* pattern (Fig. 3b) shows the reflections from two domains ("A" and "B") of rhombohedral (*R-3c*) *LSMO*, rotated by 30° in the (a,b)-plane and epitaxially grown on the substrate, i.e. $LSMO(0001)//Al_2O_3(0001)$. A coexistence of the two *LSMO* domains (A and B in Fig. 3e), low panel) is induced by the substrate itself, which allows epitaxial growth of 30°-misoriented *LSMO* domains. "*A-A*" and "*B-B*" grain boundaries (GB), with GB-angle apparently 60° are highly symmetric and can be seen only in cross-section. In contrast, the *30°-GB's* ("A-B") can be recognized in plan-view geometry (Fig. 3a)) as dark lines. In the *HRTEM* image of a triple junction in Fig. 3c) the two *30*°-GB's, i.e. "*G1-G2*" and "*G1-G3*" are clearly visible. The atomic structure of a single *30°-GB* in Fig. 3d) shows that $MnO_6$ network is not continuous across GB. We considered *30°-GB* as a quasi-periodic "zig-zag" chain of *LSMO* fragments with a lengths of *7a*. Such an interface can be rationalized within the coincidence-site lattice model[21] assuming coincidence sites as those ideally connecting the atoms across *30°-GB* (Fig. 3d). However, by superimposing of two 30°-misoriented *LSMO(0001)* planes only approximate equality, $7a_p \approx 4\sqrt{3}a_p$, holds, resulting in geometrical misfit of 1% and, hence, in strain and disorder in both *LSMO* blocks adjacent to *30°-GB*. Apparently 30°-*GB*'s, satisfying basic requirements to serve as an insulating barrier, occur very frequently (see Fig. 3a); therefore we have strong arguments to assume them to be responsible both for the tunnelling and resistance switching behavior.

As we have argued above an ionic-transport-based[7,11] mechanism of memristance cannot explain the electric-field-induced resistance switching in nano-sized LSMO/LSMO tunnel junctions as well as a re-entrant LRS-HRS switching for $U>U_C$. Thus, an alternative mechanism should account for hysteretic and bipolar resistance switching, which occurrs at a well defined voltage, $U_C \sim 8\ V$, dropping exclusively at the interfaces (GB's). Considering the number, $N=L/D$, GB's along the bridge length, $L$, the estimated critical switching field, applied to one GB, is $E_C=(1/N)*(U_C/d) \sim 10^6\ V/cm$. Very recently Maruyama et al[22] have shown that in applied electric field, $E \sim 10^6\ V/cm$, the $d$-orbitals of interfacial $Fe/MgO(100)$ will be repopulated, yielding to the change of magnetic anisotropy. Thus, we assume that JT distortion of $MnO_6$ octahedra, $\varepsilon = (d^z_{Mn-O} - d^{x,y}_{Mn-O})/d^z_{Mn-O} > 0$, in order to minimize Coulomb energy can be influenced by electric field. Indeed, in a sufficiently high external electric field the potential energy will be increased and $MnO_6$ octahedra will have no more reasons to be distorted, i.e. $\varepsilon = 0$. Such a metastable state corresponds to a maximum in the JT deformation energy, qualitatively shown in Fig. 4a). In this way the orbital degeneracy is restored and crystal symmetry is raised in applied electric field. The JT-scenario looks realistic because of the following reasons: 1) it is known[23] that interfacial (surface) *LSMO* with the thickness *L~5-7 u.c.* is orbitally reconstructed and represents itself a *CE*-type charge/orbitally ordered insulating phase[24]; 2) correlated JT polarons (*CP*) are known[25] to show the same *CE*-type architecture and their correlation length is of the same magnitude, $\delta_{JT} \sim 2$ *nm~5 u.c.* as the thickness of interfacial orbitally reconstructed region. Considering the interface as a two-dimensional defect, the *CP*'s are nucleated within the interfacial *LSMO* even without applied electric field. To actuate the switching an external field has to couple to interfacial polar moments, originated from CP's, which viewed as electric quadrupoles[18] *CP's* provide a nonlinear coupling to electric field, $N_{CP} \sim E^2$, with $N_{CP}$ is concentration of correlated polarons.

In Fig. 4b) a structural model of interfacial *LSMO* with corresponding arrangement of the $MnO_6$ octahedra is shown. At low fields, $0<U<U_C$, the *B-type* structure with *Mn-O-Mn* angle, $\alpha<180°$, and an orthorhombic ($P_{nma}$) symmetry is stable because of the static JT distortion, $\varepsilon > 0$, and enhanced stability of CE-phase at the interface. This phase corresponds naturally to *HRS* (shown in red colour in Fig. 4 c), d)). By increasing voltage the orbital repopulation (reconstruction) takes place, yielding the angle $\alpha$ to increase. By exceeding $U_C$, $\alpha$ approaches $180°$ and the structure of interfacial CE-phase transforms into a metastable phase (see a local energy maximum in Fig. 4a)) with a more symmetric, presumably *R-3c* structure, which is free of JT distortions, $\varepsilon = 0$, and of polarons. This phase can be assigned to *LRS* and is shown in green in Fig. 4b), c). By decreasing the voltage one follows a "minor loop" in Fig. 4c) due to a severe stiffness of the *R-3c* phase for $-U_C<U<U_C$. However, by further increasing the voltage, $U>U_C$, in Fig. 4d) the number of CP (quadrupoles), $N_{CP}$, will further increase but now they correspond to $\varepsilon < 0$. Finally, a re-entrant *LRS-HRS* switching takes place, bringing the system into the second energy minimum (Fig. 4a), corresponding to a stable interfacial CE-phase, presumably with orthorhombic $P_{nma}$ structure, with $\alpha<180°$. In line with the recent polaronic memristor theory of Alexandrov and Bratkovsky[14], our electronic (polaronic) memristor can be considered as a two-level vibronic MQD system with (ε>0 and ε<0) and without (ε=0) JT-distortions. Likely *30°-GB*'s (*60°-GB*'s) provide a small (large) coupling to the leads (metallic *LSMO* grains).

In summary we have observed a memristor behaviour in nano-sized LSMO/LSMO interfacial tunnel junctions due to nanocolumnar structure of LSMO/$Al_2O_3$(0001) films. The resistance switching is found to be controlled by applied electric field, which couples to the correlated polarons and influences the orbital reconstruction at the interfaces.

The work was financially supported by the DFG via SFB 602 (TP-A2) and Leibniz-Programm. O.I.L. and G.V.T. acknowledge financial support from EU FP6 (Integrated Infrastructure Initiative, Reference 026019 ESTEEM).

REFERENCES

1. S. Q. Liu, N. J. Wu and A. Ignatiev, Appl. Phys. Lett. **76**, 2749 (2000).
2. A. Beck, J.G. Bednorz, C. Gerber, C. Rossel, D. Widmer, Appl. Phys. Lett. **77**, 139 (2000).
3. M. Hamaguchi, K. Aoyama, S. Asanuma, Y. Uesu, T. Katsufuji, Appl. Phys. Lett. **88**, 142508 (2006).
4. K. Szot, W. Speier, G. Bihlmayer, R. Waser, Nature Mater. **5**, 312 (2006).
5. A. Sawa, T. Fujii, M. Kawasaki, Y. Tokura, Appl. Phys. Lett. **88**, 232112 (2006).
6. R. Waser, and M. Aono, Nature Mater. **6**, 833 (2007).
7. Y. B. Nian, J. Strozier, N. J. Wu, X. Chen, A. Ignatiev, Phys. Rev. Lett. **98,** 146403 (2007).
8. J.C. Scott, L.D. Bozano, Adv. Mater. **19**, 1452 (2007).
9. C.P. Collier, et al. Science **289**, 1172 (2000).
10. P.J. Kuekes, G.S. Snider, R.S. Williams, Sci. Am. **293**, 72 (2005) and D.B. Strukov, K.K. Likharev, J. Nanosci. Nanotechnol. **7**, 151 (2007).
11. D. B. Strukov, G. S. Snider, D. R. Stewart, R. S. Williams, Nature **453**, 80 (2008).
12. L.O. Chua, IEEE Trans. Circuit Theory **18**, 507 (1971).
13. J.J. Blackstock, W.F. Stickle, C.L. Donley, D.R. Stewart, R.S. Williams, J. Phys. Chem. C **111**, 16 (2007).
14. A.S. Alexandrov and A.M. Bratkovsky, Phys. Rev. B **80**, 115321 (2009).
15. V. Moshnyaga et al, Appl. Phys. Lett. **74**, 2842 (1999).
16. S.A. Köster et al, Appl. Phys. Lett. **81**, 1648 (2002).
17. J.A. Katine, F.J. Albert, R.A. Buhrman, E.B. Myers, D.C. Ralph, Phys. Rev. Lett. **84**, 3149 (2000).
18. V. Moshnyaga et al, Phys. Rev B **79**, 134413 (2009).
19. J.G. Simmons, J. Appl. Phys. **34**, 1793 (1963).
20. M. Bowen et al, Appl. Phys. Lett. **82**, 233 (2003).
21. N.H. Fletcher, G.E. Stillman in: J.W. Matthews (Ed.) "Epitaxial Growth, Part B", Acad. Press, N.Y. 1975, p. 529.
22. T. Maruyama et al, Nature Nanotechnology **4**, 158 (2009).
23. A. Tebano et al, Phys. Rev. Lett. **100,** 137401 (2008).
24. L. Brey, Phys. Rev. B 75, 104423 (2008).
25. C.P Adams, J.W. Lynn, Y.M. Mukovskii, A.A. Arsenov, D.A. Shulyatev, Phys. Rev. Lett. **85**, 3954 (2000).

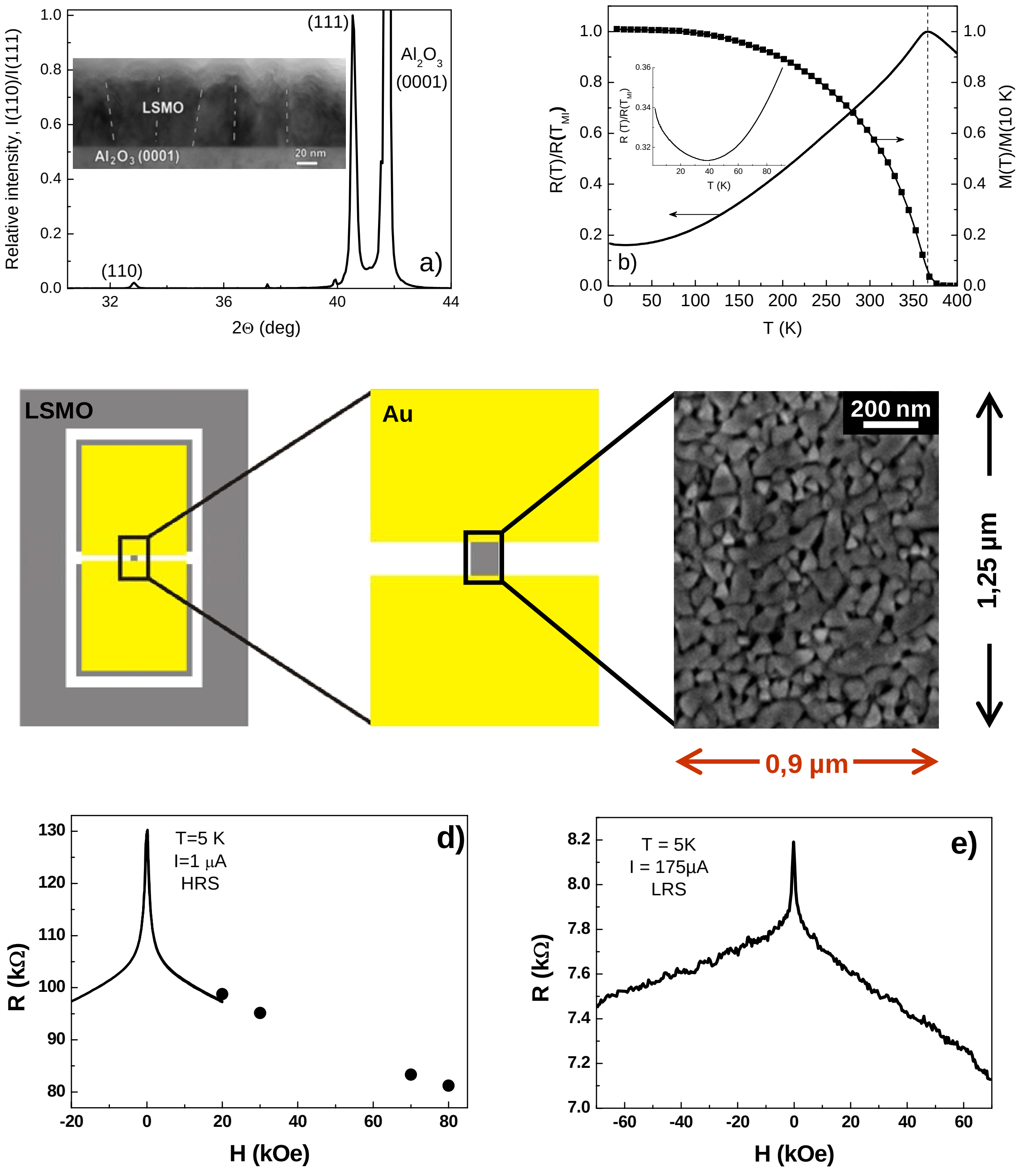

Fig. 1 a): X-ray diffraction θ-2θ pattern demonstrates a predominant (111) out-of-plan film texture. In the inset a low magnification cross section TEM image shows densely packed LSMO nanocolumns grown on $Al_2O_3$(0001); (b) normalized magnetization (right scale) and resistance both measured for the whole film. In the inset one can see low-temperature insulating behaviour for the microbridged sample; c) the geometry of a microbridge, patterned by electron beam lithography with subsequent Ar etching. Scanning electron microscopy image (right) of the microbrdige area shows triangular shaped LSMO(111) nanoblocks. The width of the bridge, W=0.9 µ, is given for the "narrow" bridge. The bottom panel shows low-field TMR for the "narrow" bridge in HRS (d) and LRS (e).

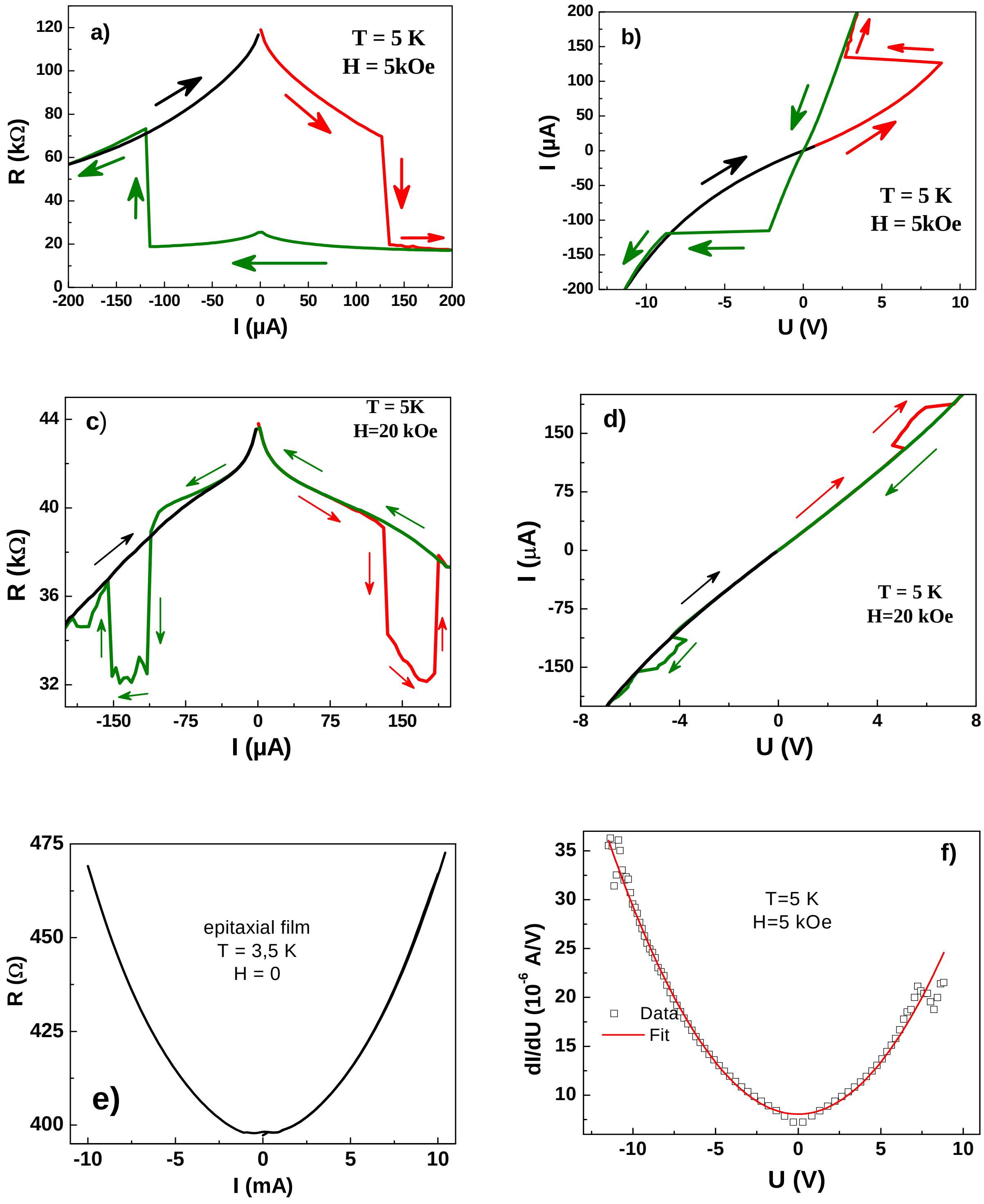

Fig. 2 The dependences of resistance on the current, R(I), (a, c) and evaluated current-voltage, I(U), characteristics (b, d) for the “narrow” (a, b) and additional (c, d) microbridges. R(I) for epitaxial microbridge (e) shows no switching effects. Differential conductivity (f) for the “narrow” microbridge illustrates tunnelling mechanism.

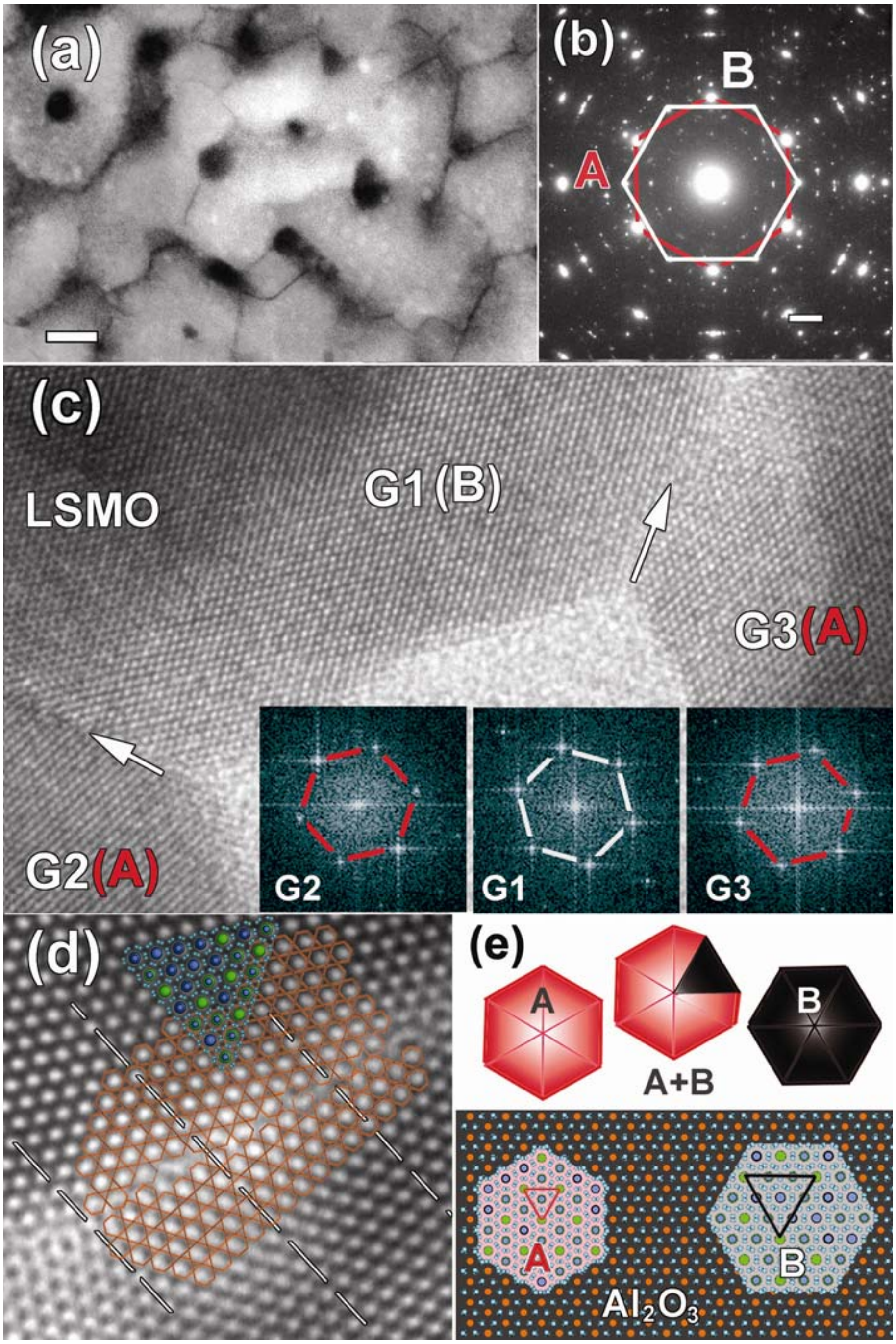

Fig. 3 (a) - High angle annular dark field plan-view STEM image of densely packed array of LSMO nanoblocks. The dark contrast lines are the 30$^o$ GB's; (b) – Corresponding ED pattern. Two orientation variants of the LSMO grains "A" and "B" are marked by red and white hexagonal frames respectively.(c) – HRTEM image of a triple junction and the corresponding FT of the different grains. Two 30$^o$ GB's are marked by white arrows; (d) – Enlargement of part of the 30$^o$GB in figure (c) together with a structural overlay. The bright dots are corresponding to the La(Sr) columns. The GB model is represent in terms of octahedra projected on the (0001) plane. The dashed lines depict the coincidence lattice sites along the 30$^o$ GB (e) Schematically presentation of the relationship between two LSMO orientation variants– A-type – red and B-type - black colored hexagons with highly symmetric 60$^o$GB with respect to (0001) $Al_2O_3$ substrate. Note that a merging of two different types (A and B) leads to the appearance of a 30$^o$ GB.

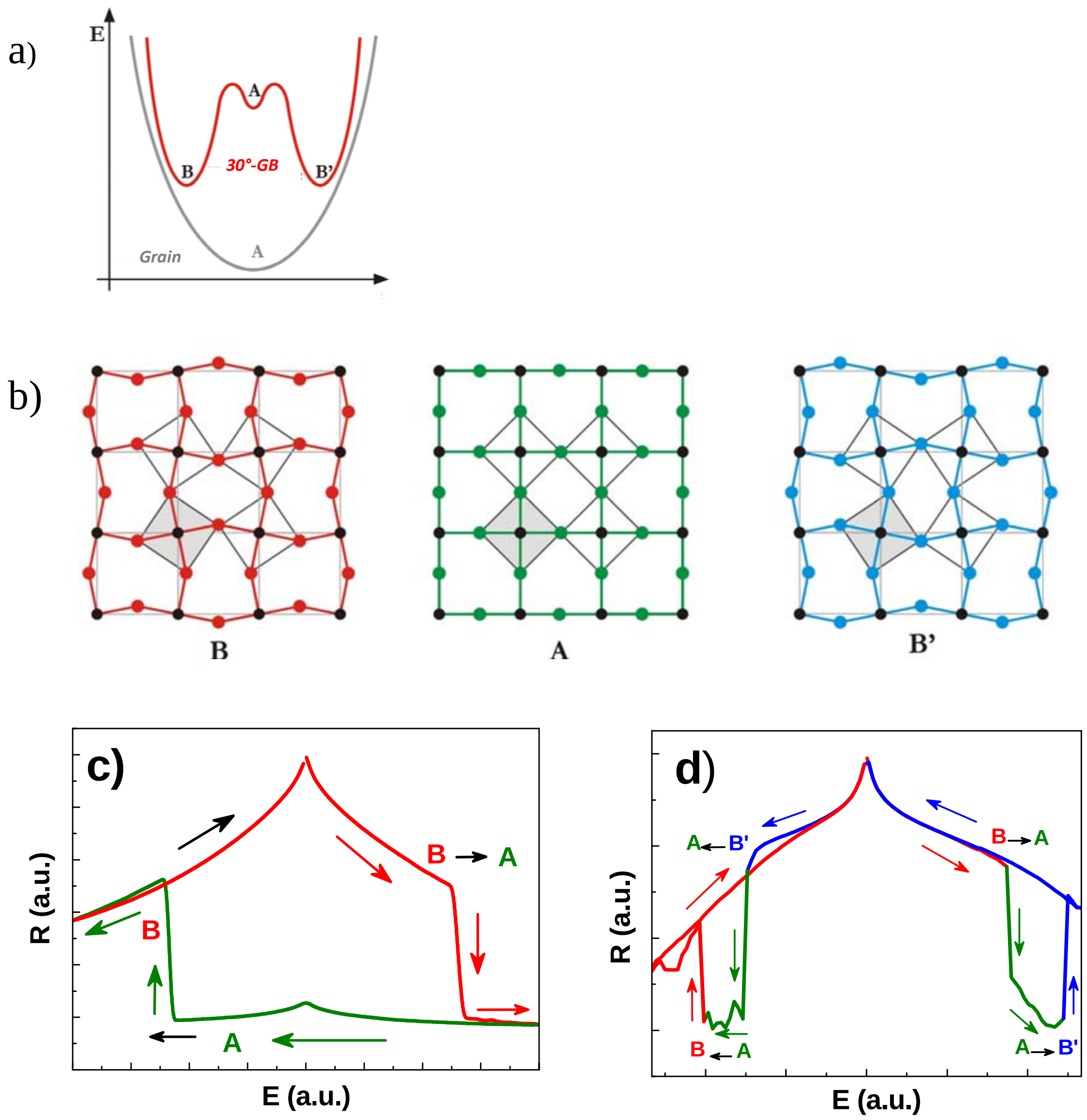

Fig. 4 Microscopic qualitative model, accounted for the existence of the two JT-energy minimums (a), associated with two (B, B') variants of $P_{nma}$ structures at the LSMO/LSMO interface, and one metastable (R-3c) structure with undistorted $MnO_6$ octahedra (A in the middle panel). In c) and d) the "minor" and "major" R(E) loops, respectively, are shown.